\documentstyle[prd,eqsecnum,preprint,tighten,aps]{revtex}
\begin{document}
 \title{High temperature asymptotics of thermodynamic functions
of electromagnetic field subjected to boundary conditions
on a sphere and cylinder}
 \draft
\author{M. Bordag\thanks{E-mail: Michael.Bordag@itp.uni-leipzig.de}}
\address{Universit\"at Leipzig, Institut f\"ur Theoretische Physik \\
Augustusplatz 10, 04109 Leipzig, Germany}
\author{V.V. Nesterenko\thanks{E-mail: nestr@thsun1.jinr.ru},
I.G. Pirozhenko\thanks{E-mail: pirozhen@thsun1.jinr.ru}}
\address{Bogoliubov Laboratory of Theoretical Physics \\ Joint
Institute for Nuclear Research, Dubna, 141980, Russia} \date{\today}
\maketitle
\begin{abstract}
The high temperature asymptotics of thermodynamic functions of
electromagnetic field subjected to boundary conditions with
spherical and cylindrical symmetries are constructed by making use
of a general expansion in terms of heat kernel coefficients and the
related determinant.  For this, some new heat kernel coefficients
and determinants had to be calculated for the boundary conditions
under consideration.  The obtained results reproduce all the
asymptotics derived by other methods in the problems at hand and
involve a few new terms in the high temperature expansions. An
obvious merit of this approach is its universality and applicability
to any boundary value problem correctly formulated.
\end{abstract}
\section{Introduction}
The Casimir effect is one of the most interesting phenomena in quantum
field theory. Since its discovery more than 50 years ago it attracted
much attention. In the past years the interest intensified after its
experimental verification reached the one percent level of precision
\cite{Mohideen,M2,M3}.

The influence of temperature on the Casimir effect was an important
topic since its first experimental demonstration \cite{Spaarnay} which
had been done at room temperature. It was first shown in Ref.\ \cite{Mehra}
that the temperature influence was just below what had been
measured. It is expected that the temperature contributions will be
seen in the upcoming series of experiments.

In quantum field theory, finite temperature effects can be described
at equilibrium in the Matsubara formalism by imposing periodic
(resp. antiperiodic for fermions) boundary conditions in the imaginary
time coordinate. Technically this is very similar to the calculation of
the Casimir effect for plane boundaries and can mathematically be
described by the same Riemann and Hurwitz zeta functions.  Another
formula important for applications is the well known Lifshitz formula
describing the interaction between plane dielectric bodies at finite
temperature.  In the case of nonflat boundaries the situation is,
however, more complicated. In order to obtain the Casimir effect at
finite temperature one has to know it at least at zero temperature,
i.e., one has to know the spectrum of the corresponding operator. Even
then complicated calculations are usually necessary and explicit
results are rare.

An opposite situation occurs with the asymptotic expansion of the
Casimir energy at high temperature. It turns out to be determined to a
large extent by local quantities which are much easier to obtain.  In
the paper \cite{DK}, which did not receive the due attention, it was
shown that this expansion can be written down in terms of the heat
kernel coefficients and the functional determinant of the operator
corresponding to the spatial part of the problem.  For example, this
method has been applied in \cite{Kir1,Kir2,Kir3} for the effective
potential in curved spacetimes and for the Casimir effect for
hypercuboids.

The heat kernel coefficients for differential operators on manifolds
with and without boundary are known to depend on the properties of
this manifold only locally. This means that they can be represented as
integrals over the manifold and over its boundary \cite{gilk95b}
whereby the characteristics like curvature enter as local
functions. The calculation of these coefficients is a topic of its own
and much progress had been achieved especially during the past decade
(see the book \cite{Avramidi2000}, for example). Less is known about
the determinant. It is not ascertain in general whether it is a local
quantity. However, in several examples it is shown to be calculable
much easier than the corresponding Casimir energy at zero
temperature. As a consequence, the high temperature expansion of the
Casimir energy can be calculated quite easily. This was emphasized in
the recent review \cite{quant-ph/0106045} where the basic formula
(2.10) is taken from.

In the present paper  we apply these general formulas to some
specific examples. First we consider parallel plates in order to
demonstrate the technical tools on a simple problem. Then we consider
the conducting spherical and cylindrical shells obtaining new terms in
the asymptotic expansion. Eventually we consider the dielectric ball
and cylinder where we restrict ourselves to the dilute
approximation. These examples demonstrate the effectiveness of the
method.

An interesting application of the general formula is the discussion of
the so called classical limit which had recently been considered in Ref.\
\cite{FMR}.  It is understood as to take place if the internal energy
which is connected with the free energy by Eq. (2.11) (see below the
next section) tends to zero for $T\to\infty$. This happens if the heat
kernel coefficients with number $n\le\frac32$ vanish.

The first calculation of the leading contributions to the high
temperature asymptotics of the Casimir energy for curved boundaries
was given in Ref.\ \cite{BD} using the multiple reflection expansion.  As it
was noticed in the recent paper \cite{Bordag:2001ta} the multiple
reflection expansion can be used for the calculation of the heat
kernel coefficients demonstrating the equivalence of both approaches
up to the question of the determinant.

The situation is to some extent different for boundaries with edges
and corners. Here the application of Riemann and Hurwitz zeta
functions seems to be more appropriate. A first example of this kind
was given in Ref.\  \cite{AW}. The appropriate more general methods can be
expected to be those given in Ref.\ \cite{eliz95b}.

The layout of the paper is as follows. In Sec.\ II the derivation of
the high temperature expansions in terms of the heat kernel
coefficients is briefly given.  In Sec.\ III the original setting of
the Casimir effect, i.e.\ parallel perfectly conducting plates in
vacuum, is considered and the high temperature asymptotics of the
thermodynamic functions are derived in terms of the relevant heat
kernel coefficients. In Sec.\ IV the high temperature asymptotics for
electromagnetic field with boundary conditions on a sphere are
obtained. In Sec.\ V the high temperature expansions are constructed
for the boundary conditions defined on the lateral of a circular
infinite cylinder.  The heat kernel coefficients needed are calculated
by making use of the respective zeta functions that have been obtained
in an explicit form in terms of the Riemann zeta function in
Ref.~\cite{LNB} and also by applying the results of Ref.\
\cite{BP}. The functional determinants entering the asymptotic
expansions at hand are calculated by making use of the technique
developed in Ref.~\cite{BGKE}. The results obtained are compared with
the high temperature asymptotics which have been derived for boundary
conditions under consideration by other methods.  The possible
extension of the approach is discussed in the Conclusions (Sec.\ VI).

The mathematical details of the calculation of the zeta determinants
are presented in Appendix A for electromagnetic field subjected to
boundary conditions given on a sphere and in Appendix B for the
boundary conditions defined on the lateral of an infinite circular
cylinder.

\section{Heat kernel coefficients and high temperature expansions}
Let the dynamics of quantum field be defined by the operator
\begin{equation}
\frac{1}{c^2}\frac{\partial^2}{\partial t^2}-\Delta,
\label{eq2_1}
\end{equation}
where $\Delta $ is not of necessity the Laplace operator, but an elliptic
differential operator depending only on space coordinates. The free
energy  $F$ of the field is determined by the  zeta function
$\zeta_{\text T}(s)$ corresponding to the Euclidean version of the
operator (\ref{eq2_1})
\begin{equation}
F=-\frac{T}{2}\,\zeta'_{\text T}(0).
\label{eq2_2}
\end{equation}
Here $T$ is the temperature measured in energy units (the Boltzmann
constant  $k_{\text B}$ is assumed to be equal to 1), and the
zeta function   $\zeta_{\text {T}}(s)$ is defined in a standard way
\begin{equation}
\zeta_{\text {T}}(s)=\sum\limits_{m=-\infty}^{\infty}\sum\limits_{\{k\}}
\left(\Omega_m^2+\omega_k^2\right)^{-s}{,}
\label{eq2_3}
\end{equation}
with
$\Omega_m=2 \pi m T/\hbar $ being the Matsubara frequencies
and $\omega_k^2/c^2$ standing for  the eigenvalues of the operator
$-\Delta $ in Eq.\ (\ref{eq2_1})
\begin{equation}
-\Delta
\varphi_k(\bbox{x})=
\frac{\omega^2_k}{c^2}\varphi_k(\bbox{x}).  \label{eq2_4}
\end{equation}

The characteristics of the quantum field system with dynamical
operator (\ref{eq2_1}) at zero
temperature are determined   by the zeta function
$\zeta (s)$ associated with the
operator $-\Delta$
\begin{equation}
\zeta(s)=\sum_{\{k\}}\omega_k^{-2 s}. \label{eq2_5}
\end{equation}
From the mathematical point of view the zeta function $\zeta(s)$
corresponding to the space part of the operator (\ref{eq2_1}) is,
undoubtedly, a simpler object  than the complete zeta function
$\zeta_{\text {T}}(s)$ because the definition (\ref{eq2_3}) involves
an additional sum over the Matsubara frequencies.
Here a natural question arises whether one can
gain knowledge of the quantum
field at nonzero temperature possessing only the zeta function $\zeta (s)$.
In Ref.\ \cite{DK} it was shown that  proceeding from the zeta function
$\zeta (s)$ one can deduce the high temperature
asymptotics of the thermodynamic functions such as Helmholtz free energy,
internal energy, and entropy.
 Let us remind briefly the derivation of these
asymptotics. By making use of the formula
\begin{equation}
\lambda^{-s}=\frac{1}{\Gamma(s)}\int_0^\infty
dt\,t^{s-1}\,e^{-\lambda t}
\label{eq2_6}
\end{equation}
the zeta  function~(\ref{eq2_3})  can be represented in the form
\begin{equation}
\zeta_{\text T}(s)=\frac{1}{\Gamma(s)}\int_0^\infty dt \,
t^{s-1}\sum_{m=-\infty}^{\infty}e^{-\Omega^2_m t}\sum_{\{k\}}^{}
e^{-\omega^2_k t}{.}
\label{eq2_7}
\end{equation}
The term with $m=0$ in this formula gives the zeta
function~(\ref{eq2_5}). In the remaining terms we  substitute
the heat kernel $K(t)$ of  the operator $- \Delta$ by its
asymptotic expansion at small~$t$
\begin{equation}
K(t)\equiv\sum_{\{k\}} e^{-\omega_k^2\,t}\simeq\frac{1}{(4 \pi t
)^{3/2}}\sum_{n=0,1/2,\dots} a_n\,t^n+\dots \,{.}
\label{eq2_8}
\end{equation}
As a result we arrive at the following asymptotic representation for the
complete zeta function $\zeta_{\text T}(s)$
\begin{equation}
\zeta_{\text T}(s)\simeq\zeta(s)+\frac{2}{(2\pi)^{3/2}}\sum_{n=0,1/2,\dots}a_n
\left(\frac{\hbar}{2 \pi T}\right)^{2 s-3+2
n}\frac{\Gamma(s-3/2+n)}{\Gamma(s)}\,\zeta_{\text {R}}(2 s+2 n-3),
\label{eq2_9}
\end{equation}
where $\zeta_{\text {R}}(s)$ is the Riemann zeta function. Taking the
derivative   of the right hand side of Eq.\ (\ref{eq2_9})
 at the point
$s=0$ and substituting the result into Eq.\ (\ref{eq2_2}) one obtains
the high temperature expansion for the free energy
\begin{eqnarray}
F(T) &\simeq&-\frac{T}{2}\zeta'(0)+a_0\frac{T^4}{\hbar^3}\,
\frac{\pi^2}{90}-a_{1/2}\,\frac{T^3}{4\pi^{3/2} \hbar^2 }
\zeta_{\text {R}}(3) -\frac{a_1}{24}\frac{T^2}{\hbar}
+\frac{a_{3/2}}{(4\pi)^{3/2}}\,T\,\ln\frac{\hbar}{T}\nonumber\\
&&-\frac{a_2}{16\pi^2}\,\hbar\, \left[\ln\left(\frac{\hbar}{4
\pi
T}\right)+\gamma\right]-\frac{a_{5/2}}{(4\pi)^{3/2}}\frac{\hbar^2}{24
T}\nonumber\\ &&-T\sum_{n\geq
3}\frac{a_n}{(4\pi)^{3/2}}\left(\frac{\hbar}{2 \pi
T}\right)^{2n-3}\,\Gamma(n-3/2)\,\zeta_{\text {R}}(2 n-3)\,{,} \quad T\to \infty {.}
 \label{eq2_10}
\end{eqnarray}
Here $\gamma$ is the Euler constant. The argument of the logarithms in
expansion (\ref{eq2_10}) are dimensional, but upon collecting similar
terms with account for the logarithmic ones in $\zeta'(0)$ it is easy
to see that finally the logarithm function has a dimensionless
argument, at least for $a_2=0$.  Let us note that according to the
definition (\ref{eq2_8}) the heat kernel coefficients in our
consideration are dimensional, because the frequencies $\omega_k$ have
the dimensionality [time]$^{-1}$.

The asymptotic expansions for the internal energy $U(T)$ and the
entropy $S(T)$ are deduced from Eq.\   (\ref{eq2_10}) employing the
thermodynamic relations
\begin{eqnarray}
U(T)&=&-T^2\frac{\partial}{\partial T}\left(T^{-1}F(T)\right),
\label{eq2_11}\\ S(T)&=&T^{-1}\left(U(T)-F(T)\right)=
 - \frac{\partial F}{\partial T}{.}
\label{eq2_12}
\end{eqnarray}
Substituting the expansion (\ref{eq2_10}) into Eqs.\  (\ref{eq2_11})
and (\ref{eq2_12}) one arrives at the asymptotics
\begin{eqnarray}
U(T)&\simeq&a_0\frac{T^4}{\hbar^3}\,\frac{\pi^2}{30}+
a_{1/2}\,\frac{T^3}{\hbar^2}
\,\frac{\zeta_{\text {R}}(3)}{2\,\pi^{3/2}}+a_1\, \frac{T^2}{24\,\hbar}\,
+\frac{a_{3/2}}{(4\pi)^{3/2}}T\nonumber\\
&&-a_2\,\frac{\hbar}{16\pi^2} \left[\ln\left(\frac{\hbar}{4\pi
T}\right)+\gamma+1\right]
-\frac{a_{5/2}}{(4\pi)^{3/2}}\,\frac{\hbar^2}{12\,T}
\nonumber\\ && -\frac{T}{4\,\pi^{3/2}}
\sum_{n\geq3}\,a_n\,\left(\frac{\hbar}{2\pi T}
\right)^{2n-3}\,\Gamma(n-1/2)\,\zeta_{\text {R}}(2n-3),
\label{eq2_13}\\
S(T)&\simeq& \frac{1}{2}\zeta'(0)+a_0\,\frac{T^3}{\hbar^3}\,
\frac{2\,\pi^2}{45}+a_{1/2}\,\frac{T^2}{\hbar^2}\,\frac{3}{4}\,
\frac{\zeta_{\text {R}}(3)}{\pi^{3/2}}
+a_1\,\frac{T}{12\,\hbar}\nonumber\\ &&
+\frac{a_{3/2}}{(4\pi)^{3/2}}\left(1-\ln\frac{\hbar}{T}\right)
-a_2\frac{\hbar}{16 \pi^2 T
}-\frac{a_{5/2}}{(4\pi)^{3/2}}\,\frac{\hbar^2}{24\, T^2
}
\nonumber\\ &&-\frac{1}{4 \,\pi^{3/2}
}\,\sum_{n\geq3}a_n\,\left(\frac{\hbar}{2\pi T}\right)^{2n-3}
 (n-2)\,\Gamma(n-3/2)\,\zeta_{\text {R}}(2n-3)\,{.}
\label{eq2_14}
\end{eqnarray}
In Eq.\ (\ref{eq2_13}) the term proportional to $a_2$ contains the logarithm
of dimensional quantity: $[\hbar/T]=[\text{time}]^{-1}$.  This is the result
of the arbitrariness arising in from the ultraviolet divergences in the case of
$a_2\not=0$ (see Ref.\ \cite{BKV} for a more detailed discussion).  Unlike this
situation, collecting the logarithm functions in the $a_{3/2}$-term and in
$\zeta'(0)$ in Eq.~(\ref{eq2_14}) leads to a dimensionless argument of the
logarithm in the final expression.

It is worth noting that the zeta determinant of the operator
$-\Delta$ (i.~e. $\zeta'(0)$) does not enter the asymptotic
expansion for the internal energy~(\ref{eq2_13}). Therefore this
high temperature expansion is completely  defined only by the heat
kernel coefficients. In view of this, the first term in the asymptotics
of the free energy in Eq.\ (\ref{eq2_10}) is referred to as a pure entropic
contribution. Its physical origin is  till now not elucidated.

\section{Perfectly conducting parallel plates in vacuum}

In this section we demonstrate the application of the high
temperature expansions (\ref{eq2_10}), (\ref{eq2_13}),
and (\ref{eq2_14}) to a simple problem of electromagnetic
field confined between
two perfectly conducting parallel plates in
vacuum. First, we   briefly recall how to construct the
zeta function in this problem.

As well known, for example, from the theory of waveguides and resonators
 \cite{HdP} the
vectors of electric and magnetic fields in the
problem at hand are expressed in terms of  the electric
($\bf{\Pi}'$) and magnetic
($\bf{\Pi}''$) Hertz vectors, each having only one
nonzero component $\Pi^{'}_z$ and $\Pi^{''}_z$ satisfying,
respectively, Dirichlet and Neumann conditions on the internal
surface of the plates. The functions $\Pi^{'}_z$ and
$\Pi^{''}_z$ obey  the equations
\begin{equation}
\left(\frac{\partial^2}{\partial
z^2}+\bbox{\nabla}^2\right)\,\Pi^{'}_z=
\frac{\omega^2}{c^2}\Pi^{'}_z,\quad
\left(\frac{\partial^2}{\partial
z^2}+\bbox{\nabla}^2\right)\,\Pi^{''}_z=
\frac{\omega^2}{c^2}\Pi^{''}_z, \label{eq3_1}
\end{equation}
where $\omega$ is the frequency of electromagnetic oscillations,
$\bbox{\nabla}^2$ stands for the two-dimensional Laplace operator for
the variables $(x,y)={\bf x}$. The separation of variables results in
the following solution
\begin{eqnarray}
\Pi^{'}_{z}({\bf x},z)&=
&\exp(i{\bf k x}
)\,\sin\left(\frac{n\pi z}{a}\right), \quad
n=1,\,2,\dots {,}\nonumber\\
\Pi^{''}_{z}({\bf x},z)&=
&\exp(i{\bf k x}
)\,\cos\left(\frac{n\pi z}{a}\right), \quad
n=0,\,1,\,2,\dots {,}\nonumber\\
\omega^2_n({\bf k})&=&
c^2\left[{\bf k}^2+
\left(\frac{n\pi}{a}\right)^2\right], \label{eq3_2}
\end{eqnarray}
where $a$ is the distance between the plates. Hence, the states of
electromagnetic field with the energy $\hbar\omega_n,\;\;n\geq 1,$ are
doubly degenerate, while the state with the energy
$\hbar\omega_0=\hbar c k$ is nondegenerate.

With allowance for this the zeta function in the problem under consideration
is given by
\begin{equation}
\zeta(s)=\frac{L_x\,L_y}{c^{2s}}\int\frac{d^2
{\bf k}}{(2\pi)^2}\left\{2\sum_{n=1}^{\infty}
\left[{\bf k}^2+\left(\frac{n\pi}{a}\right)^2
\right]^{-s}+\left( {\bf k}^2+\mu^2\right)^{-s}
\right\}, \label{eq3_3}
\end{equation}
where $L_x$ and $L_y$ are the dimensions of the plates.

For a correct definition of the integral in this formula
in the small ${\bf k}$ region
the photon mass $\mu$ is introduced (infrared regularization).
At the final step of calculations one should put $\mu=0$.  On integrating
in Eq.\ (\ref{eq3_3}) and substituting the sum over $n$ by
the Riemann zeta function one arrives at the result
\begin{equation}
\zeta(s)=\frac{L_x\,L_y}{2\pi\,c^{2s}}
\left [
\left(\frac{\pi}{a}\right)^{2-2s}\frac{\zeta_{\text {R}}(2s-2)}{s-1}+
\frac{1}{2}\frac{\mu^{2-2s}}{s-1}\right ].
\label{eq3_4}
\end{equation}

The zeta function~(\ref{eq3_4}) gives the well-known value for
the Casimir energy
\begin{equation}
E_{\text C}=\frac{\hbar}{2}\,\zeta\left(-\frac{1}{2}\right)=-c\,\hbar\,
\frac{\pi^2}{720}\,\frac{L_x L_y}{a^3} \label{eq3_6}
\end{equation}
or for its density
\begin{equation}
\frac{E_{\text C}}{V}=-\frac{c \hbar\pi^2}{720 a^4},\quad \mbox{where}
\quad V=a\,L_x \,L_y. \label{eq3_7}
\end{equation}

In order to construct the high temperature expansions (\ref{eq2_10}), (\ref{eq2_13}),
and (\ref{eq2_14})  the heat kernel
coefficients for the system under consideration should be obtained by making
use of the zeta function (\ref{eq3_4}).

The zeta function (\ref{eq3_4}) or, in the general case,
(\ref{eq2_5}) and  the
corresponding  heat kernel (\ref{eq2_8}) are related via the Mellin transform
\begin{equation}
\zeta(s)=\frac{1}{\Gamma(s)}\int_{0}^{\infty} dt\,t^{s-1}\,K(t){.}
\label{eq3_8}
\end{equation}
This enables one to  express the heat kernel coefficients $a_n$ in terms of
the values of the zeta function  at the respective points
\begin{equation}
\frac{a_n}{(4 \pi)^{3/2}}=\mathop{\lim}_{s\to\frac{3}{2}-n}
\,(s+n-3/2)\,\zeta(s)\, \Gamma(s),  \qquad n=0,\,1/2,\, \dots\,{.}
\label{eq3_9}
\end{equation}

Substituting Eq.\ (\ref{eq3_4}) into Eq.\ (\ref{eq3_9}) we obtain for
 perfectly conducting   parallel plates only one nonzero coefficient~$a_0$
\begin{equation}
a_0=2\,\frac{V}{c^2},  \label{eq3_10}
\end{equation}
where $V=L_x \,L_y\,a$ is the volume of the space  bounded by the
plates.\footnote{For obtaining the vanishing $a_{1/2}$ coefficient it is
important to take into account the second term in Eq.\ (\ref{eq3_4}) which depends
on the photon mass~$\mu$.}
This is
just an illustration of the well-known fact that for flat
manifolds without boundary or with flat boundary all the heat kernel coefficients
except for $a_0$ vanish~\cite{DeWitt}. It should be noted here
that we are considering  only  electromagnetic field confined between
the plates and
do not take into account that field outside the plates.

From Eqs.\ (\ref{eq2_13}) and (\ref{eq3_10}) it follows that the
density of internal energy  has the following high temperature
asymptotics
\begin{equation}
\frac{U(T)}{V}\simeq 4\,\frac{\sigma}{c}\,T^4, \quad T\to\infty {,}
\label{eq3_11}
\end{equation}
where $\sigma$ is the Stefan-Boltzmann constant
\begin{equation}
\sigma=\frac{\pi^2\,k_{\text B}^4}{60\,c^2\,\hbar^3}{.} \label{eq3_12}
\end{equation}
Recall that in our formulae we put $k_{\text B}=1$, that is, the
temperature is measured in energy units. The transition to
degrees is performed by the substitution $T \rightarrow k_{\text{B}}\,T$.



When calculating the high temperature asymptotics of the free
energy (\ref{eq2_10}) and the entropy (\ref{eq2_14}) one needs to
derive $\zeta'(0)$ for the zeta function (\ref{eq3_4}). Keeping in mind
that $\zeta_{\text {R}}(-2)=0$ it is convenient to use here the
Riemann reflection formula
\begin{equation}
2^{1-s}\,\Gamma(s)\,\zeta_{\text {R}}(s)\,\cos(\pi\,s/2)
=\pi^2\zeta_{\text {R}}(1-s)
\label{eq3_13}
\end{equation}
which yields
\begin{equation}
\zeta_{\text {R}}(2
s-2)\mathop{\simeq}_{s\to0}-s\,\frac{\zeta_{\text R}(3)}{2\pi^2}+{\cal
O}(s^2). \label{eq3_14}
\end{equation}
From here we deduce
\begin{equation}
\zeta'(0)=\frac{L_x\,L_y}{4\,\pi\,a^2}\,\zeta_{\text {R}}(3)=
\frac{V}{4\,\pi\,a^3}\,\zeta_{\text {R}}(3). \label{eq3_15}
\end{equation}
Insertion of Eqs.\   (\ref{eq3_10}) and (\ref{eq3_15}) into
Eq.\ (\ref{eq2_10}) gives the following high temperature behaviour
for the density of free energy
\begin{equation}
\frac{F}{V}\simeq-\frac{T}{8 \,\pi\,
a^3}\,\zeta_{\text {R}}(3)-\frac{T^4}{c^3\,\hbar^3}\,
\frac{\pi^2}{90}.\label{eq3_16}
\end{equation}
As was noted above, we are considering only  electromagnetic field
between the plates. Therefore when calculating the
Casimir forces one should drop the last term in Eq.\ (\ref{eq3_16})
since its contribution is canceled by the pressure of the black body
 radiation on the outward surfaces of the plates.
As a result the high temperature asymptotics of the
Casimir force, per unit surface area,  attracting two
perfectly conducting plates in vacuum is
\begin{equation}
{\cal F}\simeq  -\frac{T}{4 \pi a^3} \zeta_{\text {R}}(3). \label{eq3_17}
\end{equation}
Usually in the Casimir calculations the contribution of the free black body
radiation is subtracted from the very beginning~\cite{PMG}.

It is interesting  to  note that  the Casimir force (\ref{eq3_17})
and the first term on the right hand side of Eq.\ (\ref{eq3_16}) are
pure classical quantities because they do not  involve the Planck
constant $\hbar$.  These classical asymptotics seem to be derivable
without appealing to the notion of quantized electromagnetic field.
The classical limit  of the theory of the Casimir effect  is
discussed in a recent paper~\cite{FMR}.

Employing Eqs.\ (\ref{eq2_12}) and (\ref{eq3_16}) one arrives at the
high temperature  behavior of the entropy density
\begin{equation}
\frac{S(T)}{V}\simeq\frac{\zeta_{\text {R}}(3)}{8 \pi
a^3}+\frac{2\,T^3\,\pi^2}{45\,c^3\,\hbar^3}. \label{eq3_18}
\end{equation}
It is worth noting that the corrections to Eqs. (\ref{eq3_11}), (\ref{eq3_16}), (\ref{eq3_18})
are exponentially small.

The example considered shows that the zeta function of the spatial part of
evolution operator really enables one to obtain the high temperature
asymptotics of the thermodynamic functions in a straightforward way. In the
subsequent sections we shall consider quantum fields defined on manifolds with
boundaries possessing spherical or cylindrical symmetries, when the relevant
zeta functions cannot be obtained in a closed form.  Furthermore in these
cases the spectrum of the operator $-\Delta $ is not known explicitly.  And
nevertheless the method proposed is applicable to these cases also.

\section{Thermodynamic asymptotics for electromagnetic field
with boundary conditions on a  sphere}

In the present section we consider electromagnetic field
subjected to
three types of boundary conditions  on the surface
of a sphere: i) an infinitely thin and perfectly conducting  spherical shell;
ii) the surface of a sphere delimits
two material media with the same velocity of light;
iii) a dielectric ball placed in unbounded dielectric medium. In order to obtain the heat
kernel coefficients determining the high temperature asymptotics
(\ref{eq2_10}), (\ref{eq2_13}), and (\ref{eq2_14}) it is convenient to use
the explicit representation of the relevant spectral zeta functions
in terms of the
Riemann zeta function.  These formulae were derived in our recent
paper~\cite{LNB} by taking into account  the first two terms of the uniform
asymptotic expansion for the  product of the modified Bessel functions
$I_{\nu}(\nu z)\,K_{\nu}(\nu z)$.

\subsection{Perfectly conducting spherical shell}
We take advantage of Eq.\ (2.26) in Ref.\ \cite{LNB}
substituting there the variable $s$ by $2s$ and recovering the explicit dependence
on the velocity of light $c$. The latter results in the replacement of the sphere
radius by $R/c$:
\begin{eqnarray}
\zeta(s)&\simeq&\frac{1}{4}\,\left(\frac{R}{c}\right)^{2s}
s\,(1+s)\,(2+s)\left\{\left(2^{1+2s}-1\right)\,\zeta_{\text {R}}(1+2s)\,-
\,2^{1+2s}\right.\nonumber\\
&+&\left. q(s)\left[\left(2^{3+2s}-1\right)\,\zeta_{\text {R}}(3+2s)\,-
\,2^{3+2\,s}\right]+\dots\right\}, \label{eq4_1}
\end{eqnarray}
where
\begin{equation}
q(s)=\frac{1}{3840}\,(480\,+\,1736\,+\,2016\,s^2+\,568\,s^3),
\label{eq4_2}
\end{equation}
and $R$ is the radius of a sphere. The terms omitted in
Eq.\ (\ref{eq4_1}) are of the form
\begin{equation}
q_k(s) \left[\left(2^{2
(k+s)+1}-1\right)\,\zeta_{\text {R}}(2\,k+2\,s+1)-2^{2\,(k+s)+1}\right],\quad
k=2,3,4, \dots , \label{eq4_3}
\end{equation}
where $q_k(s)$  stand for some polynomials in~$s$.

Analysis of Eqs.\ (\ref{eq4_1}) and (\ref{eq4_2}) shows
that the zeta function (\ref{eq4_1})  for a perfectly
conducting spherical shell
enables one to find the exact values of the first six
heat kernel
coefficients, namely:
\begin{equation}
a_0=0, \quad a_{1/2}=0,\quad a_1=0, \quad a_{3/2}=2\,\pi^{3/2},
\quad a_2=0,\quad a_{5/2}=\frac{\pi^{3/2}}{20}\,\frac{c^2}{R^2}.
\label{eq4_4}
\end{equation}
Taking into account the structure of the  omitted terms (\ref{eq4_3}) it is
easy to see that
\begin{equation}
a_j=0,\qquad j=3,4,5,  \dots\, {.} \label{eq4_5}
\end{equation}
Having obtained the heat kernel coefficients   (\ref{eq4_4}) and (\ref{eq4_5})
we are in position to construct the high temperature asymptotics of the
internal energy of electromagnetic field by making use of Eq.~(\ref{eq2_13})
\begin{equation}
U (T)\simeq\frac{T}{4}-\left(\frac{c\,\hbar}{R}\right)^2
\frac{1}{1920\,T}+{\cal O}(T^{-3}). \label{eq4_6}
\end{equation}
Applying the technique developed in Ref.\ \cite{BKE}
more terms to this
expansion can be easily added.

In order to write the asymptotic expansions (\ref{eq2_10}) and (\ref{eq2_14})
the derivative of the zeta function at the point $s=0$ should be calculated.
Equation (\ref{eq4_1}) gives an approximate value for $\zeta'(0)$
\begin{equation}
\zeta' (0)=\frac{\gamma}{2}+\ln 2
+\frac{7}{16}\,\zeta_{\text {R}}(3)-\frac{9}{8}+\frac{1}{2}\,\ln\frac{R}{c}
=0.38265 +\frac{1}{2}\,\ln\frac{R}{c}\,{.} \label{eq4_7}
\end{equation}
The terms omitted in (\ref{eq4_1}) will render precise only the first
term in the final form of this expression, while the second term $(1/2)\,\ln(R/c)$
will not change.
The exact value of  $\zeta' (0)$ is
calculated  in  Appendix A
\begin{eqnarray}
\zeta' (0)&=&\frac{1}{2}-\frac{\gamma}{2}+\frac{7}{6}\ln
2+6\,\zeta'_{{\text R}}(-1)
+\left(-\frac{5}{8}+\frac{1}{2}\,\ln\frac{R}{c}+\ln
2+\frac{\gamma}{2} \right) \nonumber\\
&=&0.38429+\frac{1}{2}\ln\frac{R}{c}.
\label{eq4_8}
\end{eqnarray}
It is worth noting that the  expression in the round parentheses, being
multiplied by $\xi^2$,
is exactly the  value of $\zeta ' (0)$ for a compact ball with continuous
velocity
of light on its surface (see Eq.\ (\ref{eq4_16}) in the next subsection).
As a result we have the following high temperature asymptotics
of the  free energy and the entropy in the problem in question
\begin{eqnarray}
F (T)&\simeq&-\frac{T}{4}\,\left(\ln\frac{R\,T}{\hbar c
}+0.76858\right)\,-\,\left(\frac{\hbar\,c}{R}\right)^2\,
\frac{1}{3840\,T}+{\cal O }(T^{-3}), \label{eq4_9}\\
S (T)&\simeq&0.44215 +
\frac{1}{4}\,\ln\frac{R\,T}{\hbar\,c}-\frac{1}{3840}\,
\left(\frac{\hbar\,c}{R\,T}\right)^2+{\cal O}(T^{-4})\,{.}
\label{eq4_10}
\end{eqnarray}
The expression (\ref{eq4_9}) exactly reproduces the  asymptotics
obtained in Ref.\ \cite{BD} by making use of the multiple scattering
technique (see Eq.~(8.39) in that paper). We have not
calculated the coefficient $a_{7/2}$, therefore we do not know the sign
of the $T^{-3}$-correction in (\ref{eq4_9}).
In Ref.\  \cite{BD} it is noted that this term is  negative.

In Eqs.\ (\ref{eq4_6}), (\ref{eq4_9}), and (\ref{eq4_10}) the large expansion
parameter is actually a dimensionless `temperature' $\tau =RT/(\hbar c)$.
Therefore the same formulae describe the behavior of the thermodynamic functions
when $R\to \infty$ and temperature $T$ is fixed.

The high temperature asymptotics of the
thermodynamic functions  derived by making use of the general expansions
(\ref{eq2_10}), (\ref{eq2_13}), and (\ref{eq2_14})
contain the terms independent of the Planck constant
$\hbar$ or, in other words, classical contributions (see Eqs.
(\ref{eq4_6}), (\ref{eq4_9}), and (\ref{eq4_10})). This is also
true for the high temperature limit of the Casimir force calculated per
unit area  of a sphere
\begin{equation}
{\cal F} (T)\simeq-\frac{1}{4\pi R^2\,}\frac{\partial
F (T)}{\partial
R}=\frac{T}{16\pi R^3}-\left(\frac{\hbar
c}{R}\right)^2\frac{1}{4\pi R^3}\,\frac{1}{1920\,T}+{\cal
O}(T^{-3}). \label{eq4_11}
\end{equation}

The leading classical term in the asymptotics (\ref{eq4_11})
describes the Casimir force that seeks to expand the sphere. The
quantum correction in this formula stands for the Casimir pressure
exerted on the sphere surface.

    In Eqs.\ (\ref{eq4_6}), (\ref{eq4_9}) and (\ref{eq4_10}) the
Stefan-Boltzmann terms proportional to $T^4$ are absent because the
contribution of the Minkowski space was subtracted from the very
beginning in our calculations~\cite{LNB}. As a result we obtain the
vanishing heat kernel coefficient $a_0$ which, in general case, is
equal to the volume of the system under
study~\cite{quant-ph/0106045}. Therefore our results describe only
the deviation from the Stefan-Boltzmann law caused by the perfectly
conducting sphere.

The vanishing of the coefficients $a_{1/2}$ and $a_1$ in the problem
at hand can be  explained by taking into account the general properties of
the heat kernel coefficients~\cite{quant-ph/0106045} and by making
use of the results obtained in Ref.\ \cite{BKE}. As known
\cite{Stratton} the solutions to the Maxwell equations with  allowance for a
perfectly conducting sphere are expressed in
terms of the two scalar functions that satisfy the Laplace equation
with the Dirichlet and Robin boundary conditions on internal and
external surfaces of the sphere. In view of this one can write
\begin{equation}
\label{eq4_11a}
a_n=a_{n+}^{\text {D}}+  a_{n-}^{\text{D}}+ a_{n+}^{\text {R}} +
a_{n-}^{\text {R}} {,}\quad n=1/2,\,1,\, \ldots \,{,}
\end{equation}
where subscribe plus (minus) corresponds to internal
(external) region and the rest notations are obvious. In Ref.\ \cite{BKE}
it was found that
\begin{eqnarray}
\label{eq4_11b}
&a_{1/2+}^{\text {D}}=-2\pi^{3/2}R^2=
 a_{1/2-}^{\text {D}}, \quad
 a_{1/2+}^{\text {R}}= 2\pi^{3/2}R^2= a_{1/2-}^{\text {R}},&\nonumber \\
&a_{1\pm}^{\text {D}}=\pm\frac{{\displaystyle 8\pi R}}{{\displaystyle 3}}, \quad  a_{1\pm}^{\text {R}}=
\mp\frac{{\displaystyle 16 \pi R}}{{\displaystyle 3}}{.}&
\end{eqnarray}
As a result we have
\begin{equation}
\label{eq4_11c}
a_{1/2}=a_1=0\,{.}
\end{equation}

 Having
calculated the corrections to the Stefan-Boltzmann law one should
naturally discuss the possibility of their detection.  The ratio
of the leading term in
Eq.\ (\ref{eq4_6}) to the internal energy of black body radiation in
unbounded space given by the Stefan-Boltzmann law (\ref{eq3_11}) is
proportional to $\tau ^{-3}$.  Already for $\tau \sim 10$  the
corrections prove to be of order $10^{-3}$. The same value of $\tau $ can be
reached by varying the scale of length $R$ in the problem under
consideration or by respective choice of the temperature~$T$. Keeping
in mind the value of the conversion coefficient $c\hbar
=197.326$~MeV~fm $= 0.229$~K~cm \cite{RPP} we obtain the following
estimations. For $R\sim 10^{-13}$~cm (a typical hadron size) the
temperature $T$ should satisfy the inequality $T\gg 200$~MeV in order
to apply the asymptotics found.  For $R\sim 1$~cm we have $T\gg
0.229$~K and for $R\sim 7\cdot 10^{10}$~cm (radius of the Sun) the
range of applicability of the asymptotics at hand extends practically
to any temperature value $T\gg 10^{-10}$~K. Here we shall not go into
the details of a concrete experimental equipment that enables one to
observe the calculated corrections to the Stefan-Boltzmann law
confining ourselves to the estimations presented above.

\subsection{Compact ball with equal velocities of light inside
and outside}

Let us consider the spherical surface that delimits the media with
`relativistic invariant' characteristics, i.e., the velocity of light is
the same
inside and
outside the sphere~\cite{Brevik}. In this problem there naturally  arises
a dimensionless parameter~\cite{BNP}
\begin{equation}
\xi^2=\left(\frac{\varepsilon_1-
\varepsilon_2}{\varepsilon_1+\varepsilon_2}\right)^2=
\left(\frac{\mu_1-\mu_2}{\mu_1+\mu_2}\right)^2,
\label{xi}
\end{equation}
where $\varepsilon_1$ and $\varepsilon_2$ ($\mu_1$ and $\mu_2$)
are permittivities (permeabilities) inside and outside the
sphere. As usual we perform the calculation in the first order
of the expansion with respect to $\xi^2$.

In order to derive the zeta function for the  boundary
conditions at hand one should   multiply Eq.\ (\ref{eq4_1}) by $\xi^2$
and  replace there $q(s)$ by the polynomial
\begin{equation}
p(s)=-\frac{1}{2}\,\left[1-\frac{9}{2}\,(3+s)+\frac{5}{2}\,(3+s)\,
(4+s)-\frac{7}{24}\,(3+s)\,(4+s)\,(5+s)\right]{.} \label{eq4_13}
\end{equation}
The zeta function, obtained in this way, affords the exact heat kernel
coefficients up to $a_3$
\begin{eqnarray}
&a_0=0, \quad a_{1/2}=0,\quad a_1=0,\quad
a_{3/2}=2\,\pi^{3/2}\,\xi^2, \quad a_2=0,& \nonumber \\
 &\frac{{\displaystyle a_{5/2}}}{{\displaystyle (4 \pi)^{3/2}}}=
\xi ^2\frac{{\displaystyle c^2}}{{\displaystyle R^2}}\frac{{\displaystyle
p(-1)}}{{\displaystyle 8}}
=0.&\label{eq4_14}
\end{eqnarray}
With allowance of the structure of the omitted terms in Eq.\ (\ref{eq4_1})
we can again deduce that
\begin{equation}
\label{eq4_14a}
a_j=0,\quad j=3,4,5,\ldots \, {.}
\end{equation}
Substitution of  these coefficients into Eq.\ (\ref{eq2_13})
gives the  following high
temperature behavior of the internal energy in the problem under consideration
\begin{equation}
U(T)\simeq \xi^2 \,\frac{T}{4}+{\cal O}(T^{-3})\,{.} \label{eq4_15}
\end{equation}

The value of $\zeta'(0)$ is calculated in   Appendix A
\begin{equation}
\zeta'(0)=\xi^2
\left(-\frac{5}{8}+\frac{1}{2}\,\ln\frac{R}{c}+\ln
2+\frac{\gamma}{2} \right)
=\xi^2\left(0.35676+\frac{1}{2}\ln\frac{R}{c}\right).
\label{eq4_16}
\end{equation}
It is this  value that is supplied by Eq.\ (\ref{eq4_1}) after changes
specified above and with allowance for that $p(-1)=0$.

By making use of  Eqs.\ (\ref{eq2_10}), (\ref{eq4_14}), and
(\ref{eq4_16}) we deduce the high temperature asymptotics for free energy
\begin{eqnarray}
F(T)&=&-\xi^2\,\frac{T}{4}\,\left(\gamma+\ln4-\frac{5}{4}\right)+
\frac{\xi^2}{4}\,T\,\ln\frac{\hbar \,c}{R\,T}+{\cal O}(T^{-3}), \nonumber \\
&=&-\xi^2\,\frac{T}{4}0.71352+
\frac{\xi^2}{4}\,T\,\ln\frac{\hbar \,c}{R\,T}+{\cal O}(T^{-3}){.}
\label{eq4_17}
\end{eqnarray}
The entropy in the present case has the following high
temperature
behavior
\begin{eqnarray}
S(T)&=&\frac{\xi^2}{4}\,\left(1+\gamma+\ln
4-\frac{5}{4}-\ln\frac{\hbar\,c}{R\,T}\right)+{\cal O}(T^{-4}),
\nonumber\\
&=&\frac{\xi^2}{4}\,\left(1.71352-\ln\frac{\hbar\,c}{R\,T}\right)+
{\cal O}(T^{-4}).
 \label{eq4_18}
\end{eqnarray}
The asymptotics (\ref{eq4_15}) and (\ref{eq4_17}) completely
coincide with the analogous formulae obtained in Refs.\ \cite{NLS,KFMR} by
the mode summation method combined with the addition theorem for
the Bessel functions.

In Ref.\ \cite{NLS} the exact expression has also been derived for
the internal energy in the problem at hand (see Eq.\ (3.22) in that
paper). This formula gives only exponentially suppressed corrections
to the leading term (\ref{eq4_15})
\begin{equation}
\label{eq4_18a}
U(T)\simeq \xi^2\frac{T}{4}\left [
1+2(4t^2+4t+1)e^{-4t}
\right ]{,}
\end{equation}
where $t=2\pi RT$. We have used here the relation between $\xi^2$ and
$\Delta n$: $\xi ^2=\Delta n^2/4$ (see Eq.\ (3.12) in Ref.\
\cite{NLS}). The asymptotics (\ref{eq4_18a}) implies in particular
that in reality in Eq.\ (\ref{eq4_15}) there are no corrections
proportional to the inverse powers of the temperature $T$. From here
it follows immediately that all the heat kernel coefficients with
integer and half integer numbers equal or greater than 3 should vanish
\begin{equation}
\label{eq4_18b}
              a_j=0, \quad j=3,\, 7/2,\, 4,\, 5/2,\, 5,\, \ldots
\end{equation}
(compare with Eq.\ (\ref{eq4_14a})). In view of this the sign
denoting the omitted terms
in Eqs.\ (\ref{eq4_15}), (\ref{eq4_17}), and (\ref{eq4_18})
should be
substituted  by  ${\cal O} (e^{-8 \pi RT})$.

\subsection{Dielectric ball in unbounded dielectric medium}
The zeta function for electromagnetic field in the background of a
pure dielectric ball ($\mu_1=\mu_2 =1,\;\; \varepsilon_1\not=
\varepsilon _2$) has not been obtained in an explicit form. In Ref.\
\cite{BKV} the heat kernel coefficients up to $a_2$ in this problem were found.  Here
we use the results of this paper confining ourselves to the $\Delta
n^2$-approximation, where $\Delta
n=n_1-n_2=n_1\,n_2\,(c_2-c_1)/c\simeq(c_2-c_1)/c$, $n_i$ and $c_i$ are
the refractive index and the velocity of light inside ($i=1$) and
outside ($i=2$) the ball, and $c$ is the velocity of light in the
vacuum: $n_i=\sqrt{\varepsilon _i}, \quad c_i=c/n_i, \quad i=1,2$.
It is assumed that $c_1$ and $c_2$ differ from $c$ slightly,
therefore $c_2-c_1$ and $\Delta n$ are small quantities. In view of this
we have
\begin{eqnarray}
\label{eq4_19}
&a_0=\frac{{\displaystyle8}}{{\displaystyle 3}}
\pi R^3\frac{{\displaystyle c_2^3-c_1^3}}{{\displaystyle c_1^3c_2^3}}\simeq
8\pi\frac{{\displaystyle R^3}}{{\displaystyle c^3}}\,(\Delta n+2\,\Delta n^2),
&  \nonumber \\
&a_{1/2}=- 2\pi^{3/2}R^2\frac{{\displaystyle (c_1^2-c_2^2)^2}}{{\displaystyle
c_1^2c_2^2(c_1^2+c_2^2)}}
\simeq-4\,\pi^{3/2}\,\frac{{\displaystyle R^2}}{{\displaystyle c^2}}\Delta n^2,& \nonumber \\
& a_1\simeq 0,
\quad a_{3/2}=\pi^{3/2}\frac{{\displaystyle (c_1^2-c_2^2)^2}}{{\displaystyle
(c_1^2+c_2^2)^2}}
\simeq\pi^{3/2}\Delta n^2, \quad a_2\simeq 0.
&
\end{eqnarray}
The coefficients $a_1$ and $a_2$ equal zero only  in the
$\Delta n^2$-approximation considered here.
In the general case they contain
terms proportional to $\Delta n^k$, where $k\geq3$.

Allowance for one more term in  the uniform asymptotic expansion
of the modified Bessel functions, as compared with the calculations
in Ref.\ \cite{BKV}, gives the next heat kernel coefficient
\begin{equation}
\frac{a_{5/2}}{(4\pi)^{3/2}}=\frac{25}{2688}\frac{c^4}{R^2}\Delta n^4.
\label{eq4_20}
\end{equation}
Correcting the mistake made in \cite{Neapol} we state that this
coefficient has no contributions proportional to $\Delta n^2$, and
in the $\Delta n^2$-approximation one has to put
\begin{equation}
a_{5/2}\simeq 0.
\label{eq4_20a}
\end{equation}

Making use of the technique developed  in Ref.\ \cite{BGKE} one
obtains the following expression for the
derivative of the zeta function for a pure dielectric ball at the point $s=0$
(see Appendix A)
\begin{equation}
\zeta'(0)=\frac{\Delta
n^2}{4}\left(-\frac{7}{8}+\ln\,\frac{R}{c}+\ln 4 + \gamma
\right). \label{eq4_21}
\end{equation}

Before turning to the construction of the high temperature asymptotics
in the problem at hand by making use of the general formulae
(\ref{eq2_10}), (\ref{eq2_13}), and (\ref{eq2_14}) the following remark
should be done. When considering the electromagnetic field in the
background of a dielectric body in the formalism of quantum
electrodynamics of continuous media, as a matter of fact
one is dealing with a system consisting of two objects:
electromagnetic field plus a continuous dielectric body. It is
important that this body is described (phenomenologically) only by
respective permittivity without introducing into the Hamiltonian
special additional dynamical variables. As a result the zeta function
and the relevant heat kernel coefficients calculated in this formalism
also describe both electromagnetic field and dielectric body. When we
are interested in the Casimir thermodynamic functions in such
problems we have obviously to separate in the general expressions the
contributions due to the dielectric body itself~\cite{Barton2}.



Let us turn to such separation procedure in the high temperature
asymptotics for a dielectric ball.
Following the reasoning of Refs.\ \cite{Barton1,Barton}
we divide  the Helmholtz free energy of a material
body with volume $V$ and the  surface area $S$
into the parts
\begin{equation}
F=V\, f+S\,\sigma+F_{\text{Cas}}, \label{eq4.22}
\end{equation}
where $f$ is the free energy of a unit volume of a ball, $\sigma$
denotes the surface tension, and $F_{\text{Cas}}$ is refered to as the
Casimir
free energy of
electromagnetic field  connected with this body and having the
temperature $T$.
 In this way we obtain the following high
temperature behavior of the  free energy
$F(T)$ in the problem at hand
\begin{equation}
\label{eq4_23a}
F(T)\simeq a_0\frac{T^4}{\hbar ^3}\frac{\pi^2}{90}-  a_{1/2}
\frac{T^3}{4\pi^{3/2}\hbar^2}\zeta_{\text{R}}(3)+F_{\text{Cas}}(T),
\end{equation}
where $a_0$ and $a_{1/2}$ are defined in Eq.\ (\ref{eq4_19}) and
\begin{equation}
F_{\text{Cas}}(T)\simeq-\frac{\Delta n^2}{8}\,T\,\left(\ln
\frac{4\,T\,R}{\hbar\,c}+\gamma-\frac{7}{8}\right)+
{\cal O}(T^{-2}).
\label{eq4_23}
\end{equation}
The high temperature asymptotics
for the Casimir internal energy and for  the Casimir
entropy can be derived by making use of the
respective thermodynamical relations (\ref{eq2_11}), (\ref{eq2_12})
\begin{eqnarray}
U_{\text{\text{Cas}}}(T)&\simeq&\frac{\Delta n^2}{8}\,T+
{\cal O}(T^{-2}){,}
\label{eq4_24}\\
S_{\text{Cas}}(T)&\simeq&\frac{\Delta n^2}{8}\left(
\frac{1}{8}+\gamma+\ln\,\frac{4\,R\,T}{\hbar\,c}\right)+
{\cal O}(T^{-3})\,{.}
\label{eq4_25}
\end{eqnarray}

It is worth comparing  these results with
analogous asymptotics obtained by different methods.
In Ref.\ \cite{NLS}
at the beginning of calculations the first term of
expansion of internal
energy (\ref{eq4_24}) was derived. The subsequent integration of the
thermodynamic relation (\ref{eq2_11}) gave the correct
coefficient of the logarithmic term in the asymptotics of free energy
(\ref{eq4_23}).
In a very recent paper~\cite{Barton} Barton  managed to deduce the asymptotics
(\ref{eq4_23}) -- (\ref{eq4_25}). One should keep in mind
that  our parameter $\Delta n$ corresponds to
$2\pi\alpha\, n$ in the notations of Ref.~\cite{Barton}.

The asymptotics
(\ref{eq4_23})--(\ref{eq4_25}) contain  the $R$-independent terms.
As far as we know the
physical meaning of such  terms remains
unclear.

Preliminary analysis of a complete expression for the internal energy
of a dielectric ball (see Eqs.\ (3.20) and (3.31) in Ref.\
\cite{NLS}) shows that probably there are only exponentially
suppressed corrections to the leading term (\ref{eq4_24}). In that
case in addition to Eq.\ (\ref{eq4_20a}) all the heat kernel
coefficients with number greater than 3 should vanish in the
$\Delta n^2$-approximation.

\section{Thermodynamic asymptotics for  electro\-magnetic field
with boundary conditions on an infinite cylinder}

The calculation of the vacuum energy of electromagnetic field with
boundary conditions defined on a cylinder, to say nothing  of
the temperature corrections, turned out to be technically a more involved
problem than the analogous one for a sphere. Therefore the Casimir problem for
a cylinder has been considered only in a few
papers~\cite{BD,LNB,BP,DRM,MNN,NP,GoR}.
We again examine three cases: i) perfectly
conducting  cylindrical shell; ii) solid cylinder with
$c_1=c_2$; iii) dielectric cylinder when $c_1\neq c_2$.
Here we shall use the results of our previous papers \cite{LNB,BP}.
\subsection{Perfectly conducting cylindrical shell}

In Ref.~\cite{LNB} the first two terms in the uniform asymptotic expansion
of the product of the modified Bessel functions  $I_n(nx)\,K_n(nx)$ were
taken into account.   As a result
the
spectral zeta function in the problem under consideration was represented as an
expansion in
terms of the Riemann zeta functions $\zeta_{\text {R}}(2 (k+s)+1), \quad
k=0, 1,2, \ldots \;$.
With allowance for the first two terms in this expansion the zeta function
is given by
\begin{equation}
\zeta(s)=Z_1(s)+Z_2(s)+Z_3(s).
\label{eq5_1}
\end{equation}
Here the function $Z_1(s)$ stands for the contribution of
zero orbital momentum with  proper subtraction
\begin{equation}
Z_1(s)=\frac{(2 s-1)\,
R^{2s-1}}{2\,\sqrt{\pi}\,c^{2s}\,\Gamma(s)\,\Gamma(3/2-s)}\,
\int_0^{\infty}
dy\,y^{-2s}\,\left\{\ln[1-\mu_0^2(y)]+\frac{1}{4}\,y^2\,t^6(y)\right\},
\label{eq5_2}
\end{equation}
$$\mu_n(y)=y\,\left(I_n(y)\,K_n(y)\right)',\quad
t(y)=\frac{1}{\sqrt{1+y^2}}.$$
The function $Z_2(s)$ is generated by the first term of the
uniform asymptotic expansion
\begin{equation}
Z_2(s)=\frac{R^{2s-1}}{64\,\sqrt{\pi}\,c^{2s}}\,(1-2 s)\,(3-2 s)
\left[2\,\zeta_{\text {R}}(2
s+1)+1\right]\,\frac{\Gamma(1/2+s)}{\Gamma(s)}.
\label{eq5_3}
\end{equation}
The function $Z_3$ corresponds to  the  second term of the
uniform asymptotic expansion
\begin{equation}
Z_3(s)=\frac{R^{2s-1}}{61440\,\sqrt{\pi}}\,(1-2 s)\,(3-2
s)\,(784\,s^2-104\,s-235)\,\frac{\Gamma(3/2+s)}{\Gamma(s)}\,\zeta_{\text
{R}}(2s+3)\,{.}
\label{eq5_4}
\end{equation}

The function $Z_1(s)$ is defined in the strip $-3/2<\Re\, s<1/2$,
while the functions $Z_2(s)$ and $Z_3(s)$ are analytic functions  in the whole
complex plane $s$ except for the points, where  $\Gamma(s)$ and
$\zeta_{\text {R}}(s)$ have simple poles. In order to find the heat kernel
coefficients $a_0$, $a_{1/2}$, and $a_{1}$
trough the relation (\ref{eq3_9}) one needs the zeta
function defined in the region  $1/2+\varepsilon\leq \Re\, s\leq 3/2
+\varepsilon$  with $\varepsilon$ being a
positive infinitesimal. However in this region Eq.\ (\ref{eq5_2})
is not applicable directly due to the bad behaviour of the integral at the upper
limit.
In the most simple way we can overcome
this difficulty as in the case of perfectly conducting plates by
introducing the photon mass $\mu$ at the very beginning of
the calculation and making then the analytic continuation  of the zeta function
to the points $s=1/2,\,1,\,3/2$. Upon taking the residua at these
points one should  put $\mu=0$.

With regard to all this and using the relation (\ref{eq3_9}) we
find the heat kernel coefficients
\begin{equation}
a_0=0,\quad a_{1/2}=0,\quad a_1=0, \quad
 a_2=0. \label{eq5_5}
\end{equation}
The vanishing  heat kernel coefficient $a_2$  implies
that the zeta regularization gives a finite value  for the vacuum energy
in the
problem at hand~\cite{LNB,MNN}.
The coefficient  $a_{3/2} $ is determined by the function $Z_2(s)$ only (see Eq.\
(\ref{eq5_3}))
\begin{equation}
\frac{a_{3/2}}{(4\pi)^{3/2}}=\frac{3}{64\,R}.
\end{equation}
The coefficient $a_{5/2}$ is defined by the function $Z_3(s)$ given in Eq.\
(\ref{eq5_4})
\begin{equation}
\frac{a_{5/2}}{(4\pi)^{3/2}}=\frac{153}{8192}\frac{c^2}{R^3}.
\end{equation}
The calculation of the next
heat kernel coefficients  $a_3,\;a_{7/2},\dots$ would demand a
knowledge of the additional  terms in the  expansion of the
spectral zeta function
in the problem under consideration
in terms of the Riemann zeta function.
These terms are proportional to $\zeta_{\text {R}}(2 k+2 s+1)$ with
$k=2,3,\dots\,$, and may be obtained employing the technique
developed in Ref.\ \cite{LNB}. Analyzing the position of poles for these Riemann zeta
functions it is easy to show that, as well as in the spherical
case,
we have
 \[ a_j=0,\quad j=3,\,4,\,5\dots \,.\]

The zeta determinant entering the high
temperature asymptotics of  free energy (\ref{eq2_10}) and
 entropy (\ref{eq2_14}) is calculated in  Appendix B
\begin{equation}
\zeta'(0)=\frac{0.45847}{R}+\frac{3}{32\,R}\,\ln\frac{R}{2\,c}.
\label{eq5_6}
\end{equation}

Now we are able to construct the high temperature expansions of
the thermodynamic functions in the problem under consideration. For
the free energy we have
\begin{equation}
F(T)\simeq-0.22924\,\frac{T}{R}-\frac{3\,T}{64\,R}\,
\ln\frac{R\,T}{2\,\hbar\,c}-\frac{51}{65536}\,\frac{\hbar^2\,c^2}{R^3\,T}
+{\cal O}(T^{-3}).
\label{eq5_7}
\end{equation}
When comparing Eq.\  (\ref{eq5_7}) with results of other authors one
should remember that all the thermodynamic quantities that we
obtained in this section  are related to a cylinder of unit
length. The high temperature asymptotics  of the electromagnetic
free energy in presence of perfectly conducting cylindrical
shell was investigated  in Ref.\ \cite{BD}. To make the
comparison convenient we rewrite  their result as follows
\begin{equation}
F(T)\simeq-0.10362\,\frac{T}{R}-\frac{3\,T}{64\,R}\,
\ln\frac{R\,T}{2\,\hbar \,c}. \label{eq5_8}
\end{equation}
The discrepancy between the  terms linear  in $T$
in Eqs.\
(\ref{eq5_7}) and (\ref{eq5_8}) is due to the double scattering
approximation used in Ref.\ \cite{BD} (see also the next subsection). Our approach provides  an
opportunity to calculate the exact value of this term (see Eq.\  (\ref{eq5_7})).

And finally, making use of the general formulae (\ref{eq2_13})
and (\ref{eq2_14}) we derive
\begin{eqnarray}
U(T)&\simeq&\frac{3\,T}{64\,R}-\frac{153}{98304}\,\frac{c^2\,\hbar^2}{R^3\,T}+
{\cal O}(T^{-3}){,} \label{eq5_9}\\
S(T)&\simeq&\frac{0.27612}{R}+\frac{3}{64\,R}\,\ln\frac{R\,T}{2\,\hbar\,c}-
\frac{153}{196608}\,\frac{c^2\,\hbar^2}{R^3\,T^2}+{\cal O}(T^{-4}).
\label{eq5_10}
\end{eqnarray}

\subsection{Compact cylinder with $c_1=c_2$ and with $c_1\neq c_2$}
Here we consider
the boundary conditions for electromagnetic
field of two types:
i) a compact infinite cylinder with uniform velocity of light on its lateral
surface,  ii) a pure dielectric
cylinder with $c_1\neq c_2$. The explicit expressions for the
heat kernel coefficients up to $a_{2}$ we
take from Ref.~\cite{BP}, where a compact cylinder with unequal velocities of
light inside and outside was considered.
When $c_1=c_2$  the final expressions for these coefficients  are drastically
simplified
\begin{equation}
a_0=0, \quad a_{1/2}=0,\quad a_{1}=0,\quad \frac{a_{3/2}}{(4\pi)^{3/2}}
=\frac{3
\xi^2}{64\,R}, \quad a_2=0. \label{eq5_11}
\end{equation}
The zeta function   obtained for given boundary conditions  in Ref.\ \cite{LNB}
gives
\begin{equation}
\frac{a_{5/2}}{(4\pi)^{3/2}}=\xi^2\frac{c^2}{R^3}\frac{45}{8192},
\quad a_j=0,\;\;j=3,4,5, \ldots \,{.}
\label{eq5_12}
\end{equation}
The heat kernel coefficients (\ref{eq5_11}) and (\ref{eq5_12}) lead to the following
high temperature behavior of the internal energy in the problem at hand
\begin{equation}
U(T) = \frac{3\xi^2T}{64\,R}\left(
1- \frac{5}{512}\frac{c^2\hbar^2}{R^2T^2}
\right ) + {\cal O}(T^{-3})\,{.}
\label{eq5_13}
\end{equation}
The corresponding zeta determinant is calculated in Appendix B
\begin{equation}
\zeta'(0)=
\frac{\xi^2}{R}\left( 0.20699+
\frac{3}{32}\,\ln\frac{R}{2\,c}\right).
\label{eq5_14}
\end{equation}
Now we can write the high temperature asymptotics for free energy
\begin{equation}
F(T)= -\xi^2\frac{T}{R}\left [0.10350 +\frac{3}{64}\ln \frac{TR}{2\hbar c}
+\frac{15}{65536}\frac{c^2\hbar^2}{R^2T^2} \right ]
+{\cal O}(T^{-3})
\label{eq5_15}
\end{equation}
and for entropy
\begin{equation}
S(T)=\frac{\xi^2}{R} \left [
0.10350 + \frac{3}{64}\left (1+
\ln \frac{RT}{2\hbar c}
\right ) - \frac{15}{65536}\frac{c^2\hbar ^2}{T^2R^2}
\right ]+{\cal O}(T^{-4})\,{.}
\label{eq5_16}
\end{equation}

  Putting in these equations $\xi^2=1$ we arrive at the double scattering
approximation for a perfectly conducting cylindrical shell (see Eq.\
(\ref{eq5_8})). A slight distinction between the linear in $T$ terms in
Eq. (\ref{eq5_8}) and Eq.\ (\ref{eq5_15}) is due to a finite error inherent in
the numerical methods employed in both the approaches.

In the case of a pure dielectric cylinder
($\mu_1=\mu_2=1,\;\varepsilon_1\neq\varepsilon_2$)
the first four heat kernel coefficients are different from zero even in
the dilute approximation \cite{BP}
(small difference between the velocities of
light inside and outside the cylinder)
\begin{eqnarray}
a_0=-\frac{6\,\pi\,R^2}{c_2^4}\,(c_1-c_2)+
\frac{12\,\pi\,R^2}{c_2^5}\,(c_1-c_2)^2, \quad
a_{1/2}=-\frac{2\,\pi^{3/2}\,R}{c_2^4}\,(c_1-c_2)^2, \nonumber\\
a_1=\frac{8\,\pi}{c_2^2}\,(c_1-c_2)-\frac{14\,\pi}{3\,c_2^3}\,(c_1-c_2)^2,
\quad a_{3/2}=\frac{3\,\pi^{3/2}}{16 R\, c_2^2}\,(c_1-c_2)^2, \nonumber \\
 a_2=0,  \quad
 \frac{a_{5/2}}{(4\pi)^{3/2}}=
\frac{857}{61440}\frac{(c_1-c_2)^2}{R^3} \label{eq5_17}.
\end{eqnarray}
It should be noted that the coefficient $a_2$ vanishes only in the $(c_1-
c_2)^2$-approximation. As a matter of fact $a_2$ contains nonvanishing
$(c_1-c_2)^3$-terms and those of higher order~\cite{BP}. Therefore the zeta regularization
provides a finite answer for the vacuum energy of a pure dielectric cylinder
only in the $(c_1-c_2)^2$-approximation even at zero temperature.

The contribution to the asymptotic expansions
of the first three heat kernel coefficients  should be
involved into the  relevant
phenomenological parameters in the general
expression of the classical energy of a dielectric cylinder (in the same way as it
has been done for a pure dielectric ball). By making use of the coefficients
$a_{3/2}$ and $a_{5/2}$ we get the high temperature asymptotics of the internal energy in the
problem at hand
\begin{equation}
U(T)=  \Delta n^2  \frac{3}{128}\frac{T}{R}\left ( 1-
\frac{857}{17280} \frac{c^2\hbar^2}{T^2R^2}
\right ) +{\cal O}(T^{-2})\,{.}
\end{equation}
where $\Delta n=n_1-n_2\simeq(c_2-c_1)/c$.

In view of a considerable technical difficulties we shall not calculate the
zeta function determinant for a pure dielectric cylinder. We recover the respective
asymptotics of free energy by integrating the thermodynamic relation
(\ref{eq2_11}) and of entropy by using the relation (\ref{eq2_12}). Pursuing this
way we introduce a new constant of integration $\alpha $ that
remaines undetermined in our consideration
\begin{equation}
F(T)=-\Delta n^2\frac{3 }{128}\frac{T}{R}\left (\alpha + \ln
\frac{RT}{\hbar c} +\frac{857}{34560} \frac{c^2\hbar^2}{T^2R^2}
\right )+{\cal O}(T^{-2})\,{,}
\end{equation}
\begin{equation}
S(T)=\Delta n^2\frac{3}{128}
\left (1+ \alpha + \ln\frac{RT}{\hbar c}- \frac{857}{34560}
 \frac{c^2\hbar^2}{T^2R^2}
\right )+{\cal O}(T^{-3})\,{.}
\end{equation}
\section{Conclusions}

In this paper we have demonstrated efficiency and universality of the
high temperature expansions in terms of the heat kernel coefficients
for the Casimir problems with spherical and cylindrical symmetries. All
the known results in this field are reproduced  in a uniform approach
and in addition  a few new asymptotics are derived (for a compact ball
with $c_1=c_2$ and  for a pure dielectric infinite cylinder).

As the next step in the development of this approach one can try to
retain  the terms exponentially  decreasing when $T\to \infty $.
These corrections are well known, for example, for thermodynamic
functions of electromagnetic field in the presence of perfectly
conducting parallel plates \cite{PMG,LR} (see also Eq.\ (\ref{eq4_18a})).
In order to reveal  such
terms, first of all the exponentially decreasing corrections should be
retained in the asymptotic expansion (\ref{eq2_8}) for the heat kernel.

It is worth noting that in the framework of the method employed
the high temperature asymptotics can also be constructed in the problems
when the
zeta regularization does not provide a finite value of the vacuum energy at zero
temperature, i.e.\ when the heat kernel coefficient $a_2$ does not vanish.

     In Ref.\ \cite{FMR} it was argued that in the high temperature
limit  the behavior of the Casimir  thermodynamic quantities should
be the following. In the case of disjoint boundary pieces the free
energy tends to minus infinity, the entropy approaches a constant,
and the internal energy vanishes. Contributions to the Casimir
thermodynamic quantities due to each  individual connected
component  of the boundary exhibits logarithmic
deviations in temperature from the behavior just described. In our
consideration we were obviously dealing with  an individual connected
component of the boundary (a sphere or cylinder). Our
results corroborate the relevant conclusions  of Ref.\ \cite{FMR}
concerning the free energy and entropy. However the internal energy
in our calculations tends to infinity like $T$ instead to vanish,
this increase being caused by the respective logarithmic
terms in the high temperature asymptotics of  free
energy.


\acknowledgments
V.V.N.\ thanks Professor Barton for providing his
paper \cite{Barton} prior publication and for very fruitful communications.
      The work has been supported by the Heisenberg-Landau Program and by
the Russian Foundation for Basic Research (Grant No.\ 00-01-00300).
V.V.N.\ acknowledges the partial financial support of
the International Science and Technology Center  (Project No.\ 840).

\appendix
\section{Zeta function determinants for electromagnetic field subjected to
 spherically symmetric  boundary conditions}
\subsection{A perfectly conducting sphere}
First we calculate $\zeta '(0)$ (zeta determinant) for electromagnetic field in the
background of a perfectly conducting sphere.  We proceed from the
following representation for this zeta function~\cite{LNB}
\begin{equation}
\label{A1}
\zeta(s)=\left(\frac{R}{c}\right)^{2 s}\,\frac{\sin
(\pi\,s)}{\pi}\sum_{l=1}^{\infty}(2l+1)\int_0^{\infty}dy\,y^{-2s}\,
\frac{d}{d\,y}\,\ln[1-\sigma_l^2(y)]\,{,}
\end{equation}
where
\begin{equation}
\label{A2}
\sigma_l(y)=\frac{d}{d\,y}[y\, I_{\nu}(y)\,K_{\nu}(y)],
\qquad \nu=l+1/2 .
\end{equation}
 The analytic continuation of Eq.\ (\ref{A1}) to the region
$\Im \,s<0$ is performed by adding and subtracting from the integrand
its uniform asymptotics at large $\nu$
\begin{equation}
\label{A3}
\sigma^2_l(\nu \,z)\simeq\frac{t^6(z)}{4\nu^2}, \qquad
t(z)=\frac{1}{\sqrt{1+z^2}}\,{.}
\end{equation}
As a result we obtain
\begin{equation}
\label{A4}
\zeta (s)= Z(s)+ \zeta_{\text{as}}(s),
\end{equation}
where
\begin{eqnarray}
Z(s)&=&\left (\frac{R}{c}\right )^{2s}\frac{\sin (\pi s)}{2\pi}
\sum_{l=1}^{\infty} \nu^{1-2s}\int _0^{\infty} \frac{d
z}{z^{2s}}\,\frac{d}{dz
}\left\{\ln[1-\sigma_{l}^2(\nu
z)]+\frac{1}{4\,\nu^2}\,\frac{1}{(1+z^2)^3}\right\},
\label{A5}\\
\zeta_{\text{as}}(s)&=&\left (\frac{R}{c}
\right )^{2s}\frac{3\sin (\pi s)}{4\pi}\sum_{l=1}^{\infty}\nu^{-1-2s}\int_{0}^{\infty}
dz\,z^{1-2s}\,t^8(z) \nonumber \\
&=&\frac{1}{4}\left (\frac{R}{c}
\right )^{2s}s(1+s)(2+s)
\left [ (2^{1+2s}-1)\zeta_{\text R}(1+2s)-2^{1+2s}
\right ].
 \label{A6}
\end{eqnarray}

When calculating $\zeta '(0)$ one can put in Eq. (\ref{A5}) $s=0$ everywhere except
for $\sin (\pi s)$, the latter function  being substituted simply by $\pi s$.
In view of this the integral in Eq.\ (\ref{A5}) is evaluated easy if one takes into
account the limits
\begin{equation}
\label{A7}
\lim _{z\to 0}\sigma ^2_l(\nu z)=\left(
\frac{\Gamma (\nu)}{2\Gamma(\nu+1)}
\right ) ^2=\frac{1}{4\nu^2}, \quad \lim_{z\to 0} \frac{t^6(z)}{4\nu^2}=
\frac{1}{4\nu^2}
\end{equation}
and the asymptotics (\ref{A3}) at large $z$. As a result we obtain
\begin{equation}
\label{A8}
Z'(0)=-2\sum_{l=1}^{\infty}\nu \left [\ln\left (1-\frac{1}{4\nu^2}
\right )+\frac{1}{4\nu^2}
\right ]{.}
\end{equation}
Differentiation of Eq.\ (\ref{A6}) with respect to $s$ at the point $s=0$
gives
\begin{equation}
\label{A9}
\zeta '_{\text{as}}(0)=-\frac{5}{8}+\frac{1}{2}\ln R
+\ln 2+\frac{\gamma}{2}\,{.}
\end{equation}
In order to calculate  the sum over $l$ in Eq.\ (\ref{A8})
we consider  an auxiliary sum
\begin{equation}
S(a)=-\sum_{l=1}^{\infty}2\,\nu
\left [\ln\left(1-\frac{a^2}{4\nu^2}\right)+\frac{a^2}{4\,\nu^2}\right],
\quad S(0)=0,\quad S(1)=Z'(0),
\label{A10}
\end{equation}
where $a$ is a parameter.
Derivative of this sum with respect to $a$ can be rewritten in the form
\begin{equation}
\label{A11}
S'(a)=-\frac{a}{2}\sum_{l=1}^{\infty}\left [
\frac{1}{l+1/2}-\frac{1}{l+(1+a)/2}+\frac{1}{l+1/2}- \frac{1}{l+(1-a)/2}
\right ]{.}
\end{equation}
The summation in Eq.\ (\ref{A11}) can be done  by making use of the
following relations~\cite{GR}
\begin{eqnarray}
\sum_{k=1}^{\infty}\left (\frac{1}{y+k}-\frac{1}{x+k}
\right ) =\frac{1}{x}-\frac{1}{y}+\psi (x)-\psi(y), \nonumber \\
\psi (x+1) =\psi (x) +\frac{1}{x}, \quad \psi \left (\frac{1}{2}
\right )=-\gamma -2\ln 2,
\label{A12}
\end{eqnarray}
where $\psi (x)$ is the digamma function (the Euler $\psi $ function): $\psi (x)
=(d/dx)\ln \Gamma (x)$. This gives
\begin{equation}
\label{A13}
S'(a)=a\,(2-\gamma -2 \ln 2) -\frac{a}{2}
\left [ \psi \left (\frac{3}{2}+\frac{a}{2}
\right )
+ \psi \left (\frac{3}{2}-\frac{a}{2}
\right )
\right ]{.}
\end{equation}
Now we integrate the both sides of Eq.\ (\ref{A13}) over $a$
from $0$ to $1$ by making use of `Maple'
\begin{equation}
\label{A14}
S(1)=Z'(0)=  \frac{1}{2}-\frac{\gamma}{2}+\frac{7}{6}\ln 2
+6\,\zeta '_{\text{R}}(-1)\,{.}
\end{equation}
From Eqs.\ (\ref{A4}), (\ref{A9}), and (\ref{A14}) it follows that
\begin{eqnarray}
\zeta'(0)&=&\frac{1}{2}-\frac{\gamma}{2}+\frac{7}{6}\ln
2+6\,\zeta'_{\text R}(-1)+\left(-\frac{5}{8}+\frac{1}{2}\ln \frac{R}{c}+\ln
2+\frac{\gamma}{2} \right )\nonumber\\
&=& -\frac{1}{8}+\frac{13}{6}\ln 2+6\zeta'_{\text{R}}(-1)+
\frac{1}{2}\ln \frac{R}{c} \nonumber \\
&=&0.38429+\frac{1}{2}\ln \frac{R}{c}. \label{A999}
\end{eqnarray}

\subsection{A material ball with $c_1=c_2$}

The same technique can be used for calculating the  zeta function
determinant in the case of equal velocities of light inside and
outside the material ball (see Section IV.B). The complete zeta function in this problem
has the form~\cite{LNB}
\begin{equation}
\zeta(s)=\left(\frac{R}{c}\right)^{2 s}\,\frac{\sin
(\pi\,s)}{\pi}\sum_{l=1}^{\infty}(2
l+1)\,\int_0^{\infty}\,dy\,y^{-2s}\,
\frac{d}{d\,y}\,\ln[1-\xi^2\sigma_l^2(y)], \label{A16}
\end{equation}
where  $\sigma _l(y)$ is defined in Eq.\ (\ref{A2}) and the parameter $\xi ^2$
was introduced in Eq.\ (\ref{xi}).
Adding and subtracting under the integral sign in Eq.\ (\ref{A16}) the
uniform  asymptotics of the integrand
at large $\nu$ we get
\begin{eqnarray}
\zeta(s)&=&\left (\frac{R}{c}\right )^{2s}\frac{\sin(\pi s)}{2\pi}
\sum_{l=1}^{\infty} \nu^{1-2s}\int_0^{\infty} \frac{d z}{z^{2s}}
\,\frac{d}{d z
}\left\{\ln[1-\xi^2\,\sigma_{l}^2(\nu
z)]+\frac{\xi^2}{4\,\nu^2}\,\frac{1}{(1+z^2)^3}\right\}\nonumber\\
&&+ \xi^2\zeta_{\text{as}}(s),
 \label{A17}
\end{eqnarray}
where the function $\zeta_{\text{as}}(s)$ was introduced in Eq.\ (\ref{A6}).
Proceeding in the same  way as in the previous subsection
we obtain for the derivative of the function $\zeta (s)$ at the point $s=0$
\begin{equation}
\zeta'(0)=S(\xi^2) +\xi ^2 \zeta'_{\text{as}}(0),
 \label{A18}
\end{equation}
where the function $S(\xi ^2)$ is defined in Eq.\ (\ref{A10}).
For small values of the argument $\xi^2$  we deduce from Eq.\ (\ref{A10})
\begin{equation}
\label{19}
S(\xi^2)=\frac{\xi^4}{16}\sum_{l=1}^{\infty}\frac{1}{\nu ^3}=
\frac{\xi^4}{16}\left [7\,\zeta_{\text{R}}(3)-8\right ]+{\cal O}(\xi^6).
\end{equation}
Therefore restricting ourselves to the first order of
$\xi^2$ we arrive at the final result
\begin{equation}
\zeta'(0)=\xi^2\,\zeta'_{\text{as}}(0)
=\xi^2\,\left(-\frac{5}{8}+\frac{1}{2}\ln
\frac{R}{c}+\ln 2+\frac{\gamma}{2} \right ). \label{A20}
\end{equation}

\subsection{A pure dielectric ball}
A material ball with arbitrary velocities of light inside
and outside treated in  Section IV.C proves to be a more complicated
problem. In
notations of Ref.\ \cite{BKV} the relevant zeta function takes the
form $\zeta(s)=\zeta_{-1}(s)+\zeta_{1}(s)$, where
\begin{equation}
\zeta_{\rho}(s)=-\frac{2\, R^{2s}}{\Gamma(s+1)\,\Gamma(-s)}
\sum_{l=1}^{\infty}\nu^{1-2s}\int_0^{\infty}dk\,k^{-2s}\frac{d}{d
k}\ln\Delta_{\rho,l}(\nu k), \quad \nu=l+\frac{1}{2}, \;\;
\rho=\pm1 \label{}
\end{equation}
with
\begin{equation}
\Delta_{\rho,l}(\nu
k)=\frac{2\,e^{-(k_1-k_2)\,\nu}}{(\chi^{\rho}+1)}
\left[\chi^{\rho}s'_l(\nu k_1)\,e_l(\nu k_2)-s_l(\nu
k_1)\,e'_l(\nu k_2)\right], \quad k_{1}=k/c_{1},\;\;k_2=k/c_2,
\label{A22}
\end{equation}
\[
 s_l(y)=\sqrt{\frac{\pi y}{2}}\,I_{\nu}(y),\quad
e_l(y)=\sqrt{\frac{2\,y}{\pi}}\,K_{\nu}(y).
\]
The parameter
$\chi=\sqrt{(\varepsilon_1\,\mu_2)/(\varepsilon_2\,\mu_1)}$
corresponds to $\xi$ in Ref.\ \cite{BKV}.

The analytic continuation of the zeta function at hand to the region
$\Re\, s>0$ is performed by adding and subtracting in
Eq.\ (\ref{A22})
several terms of its asymptotic expansion
\begin{equation}
\Delta_{\rho,l}(\nu k)\sim
\sum_{n=-1,0,1}^{\infty}\frac{D_{n,\rho}}{\nu^n}, \label{A23}
\end{equation}
where
\begin{eqnarray}
D_{-1}&=&\eta(k_1)-\eta(k_2)-(k_1-k_2), \quad
\eta(z)=\sqrt{1+z^2}+\ln\frac{z}{1+\sqrt{1+z^2}}, \nonumber\\
D_0&=&\ln\left\{\frac{\chi^{\rho}\,c_1\,t_2+
c_2\,t_1}{\sqrt{c_1\,c_2\,t_1\,t_2}\,(\chi^{\rho}+1)}\right\},
\quad t_i=\frac{1}{\sqrt{1+k_i^2}},\;\;i=1,\;2. \label{A24}
\end{eqnarray}
For our purpose it is sufficient to consider four leading terms
of the asymptotic expansion~(\ref{A23}),  $n=-1,\,0,\,1,\,2.$
The functions $D_{1}$ and $D_{2}$ are given in Ref.\ \cite{BKV}.
Proceeding in this way we represent the zeta function (\ref{})
as follows
\begin{eqnarray}
\lefteqn{\zeta_{\rho}(s)
=-\frac{2\,R^{2s}}{\Gamma(s+1)\,\Gamma(-s)}\,
\sum_{l=1}^{\infty}\,\nu^{-2s+1}\int _0^{\infty}dk\,k^{-2s}\,\frac{d}{d
k}\,\left (\ln\Delta_{\rho,l}-\nu\,D_{-1}-D_0-\frac{D_1}{\nu}-\frac{D_2}{\nu^2}\right )
}\nonumber\\ &&-\frac{2\,R^{2s}}{\Gamma(s+1)\,\Gamma(-s)}
\left[\zeta_{\text{H}}\left(2s-2,\frac{3}{2}\right)\,
\int _0^{\infty}dk\,k^{-2s}\, \frac{d
D_{-1}}{d  k}+\zeta_{\text{H}}\left(2s-1,\frac{3}{2}\right)\,
\int_0^{\infty}dk\,k^{-2s}\, \frac{d  D_0}{d
k} \right.\nonumber\\
&&\left.+\zeta_{\text{H}}\left(2s,\frac{3}{2}\right)\,
\int _0^{\infty}dk\,k^{-2s}\, \frac{d  D_1}{d
k} +\zeta_{\text{H}}\left(2s+1,\frac{3}{2}\right)\,
\int_0^{\infty}dk\,k^{-2s}\, \frac{d  D_2}{d
k} \right ], \label{A25}
\end{eqnarray}
where $\zeta_{\text{H}}$ is the Hurwitz zeta function.
Taking the derivative of the zeta function (\ref{A25}) at the
point $s=0$ with allowance for the behavior of $D_i(k)$ at $k=0$
and $k=\infty$ we obtain
\begin{eqnarray}
\zeta'_{\rho}(0)&=&-2\sum_{l=1}^{\infty}\left(l+\frac{1}{2}\right)
\left[\ln\left(1+\frac{1}{2\nu}\,
\frac{\chi^{\rho}\,c_1-c_2}{\chi^{\rho}\,c_1+c_2}\right)-
\frac{1}{2\nu}\,\frac{\chi^{\rho}\,c_1-c_2}{\chi^{\rho}\,c_1+c_2}
+\frac{1}{8\nu^2}\,\left(\frac{\chi^{\rho}\,c_1-
c_2}{\chi^{\rho}\,c_1+c_2}\right)^2\right ]\nonumber\\
&&+2\left\{\frac{1}{4}\,\ln\frac{c_2}{c_1}+\frac{11}{24}\,
\ln\left[\frac{\chi^{\rho}\,c_1+c_2}{\sqrt{c_1\,c_2}\,
(\chi^{\rho}+1)}\right]+\frac{1}{2}\,\frac{\chi^{\rho}\,c_1-
c_2}{\chi^{\rho}\,c_1+c_2}\right.\nonumber\\
&&\left.-\frac{1}{8}\,(2-\ln R-\gamma-2\,\ln
2)\,\left(\frac{\chi^{\rho}\,c_1-
c_2}{\chi^{\rho}\,c_1+c_2}\right)^2-\int _0^{\infty}dk\,\ln
k \frac{d }{d  k}\,D_2 \right\}{.} \label{A26}
\end{eqnarray}
In order to calculate the sum over $l$ in Eq.\ (\ref{A26}) we
consider an auxiliary sum
\begin{equation}
S_1(b)=\sum_{l=1}^{\infty}2\,\nu
\left[\ln\left(1+\frac{b}{2\,\nu}\right)-\frac{b}{2\,\nu}+
\frac{b^2}{8\,\nu^2}\right], \quad S_1(0)=0\,{,}\label{A27}
\end{equation}
with $b$ being a parameter.
The derivative of this sum can be cast to the form
\begin{equation}
\label{A28}
S'(b)=\frac{b}{2}\sum_{l=1}^{\infty}\left [
\frac{1}{l+1/2}-\frac{1}{l+(b+1)/2}
\right ]{.}
\end{equation}
Taking into account Eqs.\ (\ref{A12}) we obtain
\begin{equation}
S'(b)=
 \frac{b}{2}\,(-2+\gamma+2\ln
2)+\frac{b}{2}\psi\left(\frac{b}{2}+\frac{3}{2}\right).
\label{A29}
\end{equation}
The integration of Eq.\  (\ref{A29}) over $b$
from $0$ to $b=(\chi^{\rho}\,c_1-c_2)/(\chi^{\rho}\,c_1+c_2)$
gives the sum entering Eq. (\ref{A26})
\begin{eqnarray}
\lefteqn{S_1(b)
=\frac{b^2}{4}\,(-2+\gamma+2\,\ln
2)+b\,\zeta_{\text{H}}'\left(0,\frac{3}{2}+
\frac{b}{2}\right)}\nonumber\\
&&-2\left[\zeta_{\text{H}}\left(-1,\frac{3}{2}+\frac{b}{2}\right)+
\zeta_{\text{H}}'\left(-1,\frac{3}{2}+\frac{b}{2}\right)\right]
+2\left[\zeta_{\text{H}}\left(-1,\frac{3}{2}\right)+
\zeta_{\text{H}}'\left(-1,\frac{3}{2}\right)\right]{.} \label{A30}
\end{eqnarray}
Substitution of Eq.\ (\ref{A30}) into Eq.\  (\ref{A26}) gives
\begin{eqnarray}
\zeta'_{\rho}(0)&=& -\frac{b^2}{4}\,(-2+\gamma+2\,\ln
2)-b\,\zeta_{\text{H}}'\left(0,\frac{3}{2}+ \frac{b}{2}\right)\nonumber\\
&&+2\left[\zeta_{\text{H}}\left(-1,\frac{3}{2}+\frac{b}{2}\right)+
\zeta_{\text{H}}'\left(-1,\frac{3}{2}+\frac{b}{2}\right)\right]
-2\left[\zeta_{\text{H}}\left(-1,\frac{3}{2}\right)+
\zeta_{\text{H}}'\left(-1,\frac{3}{2}\right)\right]\nonumber\\
&&+2\left\{\frac{1}{4}\,\ln\frac{c_2}{c_1}+\frac{11}{24}\,
\ln\left[\frac{\chi^{\rho}\,c_1+c_2}{\sqrt{c_1\,c_2}\,
(\chi^{\rho}+1)}\right]+\frac{b}{2}-\frac{b^2}{8}\,(2-\ln
R-\gamma-2\,\ln 2)\right.\nonumber\\
&&\left.-\int _0^{\infty}dk\,\ln k
\frac{d }{d  k}\,D_2 \right\}, \qquad
b=\frac{\chi^{\rho}\,c_1-c_2}{\chi^{\rho}\,c_1+c_2}. \label{A31}
\end{eqnarray}

In the case of nonmagnetic media ($\mu_1=\mu_2=1$) the
right-hand side of Eq.\  (\ref{A31}) is slightly simplified.
Assuming that we are dealing with a dilute dielectric ball we can
expand $\zeta '(0)$ in powers of the difference
$(c_1-c_2)$, where
$c_1=1/\sqrt{\varepsilon_1},\;\;c_2=1/\sqrt{\varepsilon_2}$.
As a result we get
\begin{equation}
\zeta'(0)=\zeta'_{\rho=-1}(0)+\zeta'_{\rho=1}(0)=\frac{1}{4}
\left(-\frac{7}{8}+\ln \frac{R}{c_2}+\ln
4+\gamma\right)\frac{(c_1-c_2)^2}{c_2^2}+{\cal O}((c_1-c_2)^3). \label{A32}
\end{equation}

\section{Zeta function determinants for electromagnetic field with
 cylindrically symmetric  boundary conditions}
\subsection{A perfectly conducting cylindrical shell}
A complete spectral zeta function in the
problem at hand is defined by the following expression~\cite{LNB}
\begin{eqnarray}
\zeta(s)&=&\frac{R^{2s-1}}{2\sqrt{\pi}c^{2s}
\Gamma(s)\,\Gamma(3/2-s)}\int _0^{\infty}dy\,y^{1-2s}
\frac{d}{dy}\ln[1-\mu _0^2(y)]\nonumber\\
&&+\frac{R^{2s-1}}{\sqrt{\pi}c^{2s}
\Gamma(s)\,\Gamma(3/2-s)}\sum_{n=1}^{\infty}n^{1-2s}
\int _0^{\infty}dy\,y^{1-2s}
\frac{d}{d\,y}\,\ln[1-\mu _n^2(ny)],
\label{B1}
\end{eqnarray}
where
\[
\mu _n(y)=y\frac{d}{dy}[I_{n}(y)K_{n}(y)].
\]
The first term on the right hand side of Eq.\ (\ref{B1}) is an analytic function
of the complex variable $s$ in the strip $-1/2< \Re\, s < 1/2$.
Therefore there is no need in analytic continuation of this expression
when calculating $\zeta '(0)$. As regard to the second term  in Eq.\ (\ref{B1})
its analytic continuation to the region $\Re\, s<0$ can be accomplished in a
standard way. We add and subtract here the uniform asymptotics of the
integrand when $n$ tends to infinity
\begin{equation}
\ln[1-\mu ^2_n(ny)]\simeq-\frac{y^4\,t^6(y)}{4\,n^2}+
{\cal O}(n^{-4}),\quad t(y)=\frac{1}{\sqrt{1+y^2}}.
\label{B2}
\end{equation}
As a result we obtain
\begin{eqnarray}
\zeta(s)&=&\frac{R^{2s-1}}{2\sqrt{\pi}\,c^{2s}
\Gamma(s)\,\Gamma(3/2-s)} \int_0^{\infty} \frac{d
y}{y^{2s-1}}\frac{d }{d  y
}\ln[1-\mu ^2_{0}(y)]\nonumber\\
&&+\frac{R^{2s-1}}{\sqrt{\pi}c^{2s}
\Gamma(s)\,\Gamma(3/2-s)}\sum_{n=1}^{\infty}n^{1-2s}
\int _0^{\infty} \frac{d
y}{y^{2s-1}}\frac{d }{d  y
}\left\{\ln[1-\mu ^2_{n}(ny)]+\frac{y^4\,t^6}{4\,n^2}\right\}
\nonumber\\&&-\frac{R^{2s-1}}{32\sqrt{\pi}c^{2s}}
(1-2s)(3-2 s)\zeta_{\text{R}}(2s+1) \frac{\Gamma(1/2+s)}{\Gamma(s)}.
\label{B3}
\end{eqnarray}
Keeping in mind the behavior of the gamma  function at the  origin
$\Gamma (s)\simeq s^{-1}$  one can easily find the derivative of $\zeta (s)$
at the point $s=0$
\begin{eqnarray}
\lefteqn{\zeta'(0)=
\frac{1}{\pi R}\int _0^{\infty}dy\,y\frac{d}{dy}
\ln[1-\mu _0^2(y)] }\nonumber  \\
&&+\frac{2}{\pi\,R}\sum_{n=1}^{\infty}n
\int _0^{\infty}dy\,y\frac{d}{dy}
\left\{\ln[1-\mu _n^2(n\,y)]+\frac{y^4 t^6}{4n^2}\right\}
+\frac{1}{32R}\,\left(3\gamma-4-3\ln\frac{2\,c}{R}\right).
\label{B4}
\end{eqnarray}
Unlike the spherically symmetric boundaries,  the integration is not removed
in the formula obtained for $\zeta '(0)$.
Therefore the first
two terms in Eq.\ (\ref{B4}) can be  calculated only  numerically
\begin{equation}
\label{B5}
-\frac{1}{\pi R}\int _0^{\infty}dy
\ln[1-\mu _0^2(y)] =\frac{0.53490}{R}{.}
\end{equation}
Applying the FORTRAN subroutine that approximates the Bessel functions by
Chebyshev's polynomials we evaluate the first 30 terms in the sum
in Eq.\ (\ref{B4})
\begin{equation}
\label{B6}
-\frac{2}{\pi\,R}\sum_{n=1}^{\infty}n
\int _0^{\infty}dy
\left\{\ln[1-\mu _n^2(n\,y)]+\frac{y^4 t^6}{4n^2}\right\}=-\frac{0.00554}{R}{.}
\end{equation}
Finally  gathering together all these results we have
\begin{equation}
\zeta'(0)=\frac{0.45847}{R}+\frac{3}{32\,R}\,\ln\frac{R}{2\,c}.
\label{B7}
\end{equation}

\subsection{A compact infinite cylinder with $c_1=c_2$}

Now we turn to a compact cylinder placed into
unbounded medium such that the velocity of light is uniform
on the lateral  surface of the cylinder. Proceeding
as in the case of a cylindrical shell we start with  the expression
for the relevant spectral zeta function
\begin{equation}
\zeta(s)= \frac{R^{2s-1}}{2\,\sqrt{\pi}c^{2s}
\Gamma(s)\,\Gamma(3/2-s)}\sum_{n=-\infty}^{\infty}\,
\int _0^{\infty}\,dy\,y^{1-2s}\,
\frac{d}{dy}\ln[1-\xi^2\mu _n^2(y)] \label{B8}
\end{equation}
with parameter $\xi$ determined in Eq.\ (\ref{xi}).
In the linear approximation with respect to $\xi^2$ Eq.\ (\ref{B8}) assumes
the form
\begin{eqnarray}
\zeta(s)&=&
-\frac{R^{2s-1}\,\xi^2}{2\sqrt{\pi}c^{2s}
\Gamma(s)\Gamma(3/2-s)}\int_0^{\infty}dy\,y^{1-2s}
\frac{d}{dy}\mu _0^2(y)\nonumber\\
&&-\frac{R^{2s-1}\,\xi^2}{\sqrt{\pi}c^{2s}
\Gamma(s)\Gamma(3/2-s)}\sum_{n=1}^{\infty}
\int _0^{\infty}dy\,y^{1-2s}
\frac{d}{dy}\mu _n^2(y). \label{B9}
\end{eqnarray}
The analytic continuation to the region $\Re \,s<0$ is
needed only for the second term in Eq.\ (\ref{B9}).
Adding and subtracting here the uniform asymptotics of the integrand
for large~$n$
\begin{equation}
-\mu ^2_n(ny)\simeq-\frac{y^4t^6(y)}{4n^2}
+{\cal O}(n^{-4}), \label{B10}
\end{equation}
we obtain
\begin{eqnarray}
\zeta'(0)&=&
-\frac{\xi^2}{\pi\,R}\int _0^{\infty}dy\,y\frac{d}{dy}
\mu _0^2(y)+\frac{2\,\xi^2}{\pi\,R}\sum_{n=1}^{\infty}n
\int _0^{\infty}dy\,y\,\frac{d}{dy}\,
\left[-\mu _n^2(n\,y))+\frac{y^4\,t^6}{4\,n^2}\right]\nonumber\\
&&+\frac{\xi^2}{32\,R}\,\left(3\gamma-4-3\ln\frac{2\,c}{R} \right) {.}
\label{B11}
\end{eqnarray}
The first two terms in Eq.\ (\ref{B11})
can again be calculated only numerically
\begin{equation}
\label{B12}
\frac{\xi^2}{\pi R}\int_{0}^{\infty}dy\, \mu_0^2(y)=\frac{\xi^2}{R}0.28428{,}
\end{equation}
\begin{equation}
\label{B13}
-\frac{2\xi^2}{\pi R}\sum_{n=1}^{\infty}n\int_{0}^{\infty}dy
\left [ -\mu^2_n(ny)+\frac{y^2t^6}{4n^2}
\right ]
=-\frac{\xi^2}{R}0.00640{.}
\end{equation}
The final result reads
\begin{eqnarray}
\zeta'(0)&=&\frac{\xi^2}{R}\left[0.28428-0.00640+
\frac{1}{32}\left(3\gamma-4-3\ln\frac{2\,c}{R}\right)\right]\nonumber\\
&=& \frac{\xi^2}{R}\left(0.20699+
\frac{3}{32}\ln\frac{R}{2c}\right). \label{B14}
\end{eqnarray}


\end{document}